\documentclass[onecolumn,nobibnotes,nofootinbib]{revtex4}
\usepackage{amsmath,amssymb}
\usepackage{graphicx,subfigure,epsfig}

\newcommand{\be}{\begin{equation}}
\newcommand{\ee}{\end{equation}}
\newcommand{\bea}{\begin{eqnarray}}
\newcommand{\eea}{\end{eqnarray}}
\newcommand{\benn}{\begin{displaymath}}
\newcommand{\eenn}{\end{displaymath}}
\newcommand{\beann}{\begin{eqnarray*}}
\newcommand{\eeann}{\end{eqnarray*}}
\newcommand{\oone}{
\begin{picture}(10,8)
\put(5,5){\circle{8}}
\put(2.9,2.5){{\scriptsize 1}}
\end{picture}
}

\begin{document}

\title{High energy scattering in the saturation regime including running coupling and rare fluctuation effects}
\author{Wenchang Xiang\footnote{wcxiang@physik.uni-bielefeld.de}}
\affiliation{Fakult$\ddot{a}$t f$\ddot{u}$r Physik, Universit$\ddot{a}$t Bielefeld, D-33501 Bielefeld, Germany}
\date{\today}

\begin{abstract}
The analytic result for the $S$-matrix in the saturation regime including the running coupling is obtained. To get this result
we solve the Balitsky and Kovchegov-Weigert evolution equations in the saturation regime, which include running coupling corrections.
We study also the effect of rare fluctuations on top of the running coupling. We find that the rare fluctuations are less important
in the running coupling case as compared to the fixed coupling case.
\end{abstract}

\maketitle


\section{Introduction}
\label{sec:intro}
The Balitsky-Kovchegov (BK) equation~\cite{B,K} is a non-linear evolution equation which
describes the high energy scattering of a q$\bar{q}$ dipole on a target in the case of fixed
coupling. An analytic solution to the BK equation in the saturation region has been found by
Levin and Tuchin~\cite{LT}. The BK equation can be viewed as a mean field version of more complete equation~\cite{B}
where the higher correlations are neglected: The $S$-matrix of the
scattering of two QCD dipoles on a target is replaced in the BK equation by the product of the $S$-matrices of the individual
dipoles. Such a replacement is legitimate only in the absence of fluctuations in the light cone
wavefunction of the target~\cite{SH}. However, in Ref.~\cite{EM} was shown that rare fluctuations do change the result for
the $S$-matrix in the saturation region.

Recently, the evolution equations which include running coupling
effects have been derived by Balitsky and
Kovchegov-Weigert~\cite{Bnlo,KW}. They found that the running
coupling corrections are included in the BK kernel by replacing
the fixed coupling $\alpha_s$ in it with a ``triumvirate'' of the
running couplings. A more complete evolution equation has been
studied by Albacete and Kovchegov~\cite{JA}, they have calculated
in addition to the Balitsky and Kovchegov-Weigert equations also
the so called subtraction contributions. A numerical solution of
the more complete evolution equations were given in~\cite{JA}.

In this work, we will analytically solve these equations in the
saturation region and obtain an analytic
result for the $S$-matrix. We find that the running coupling
corrections modify the $S$-matrix a lot as compared to the fixed
coupling case. Moreover, we study the effect of the rare
fluctuations on top of the running coupling in the way as it was
done in Ref.~\cite{EM} for the fixed coupling case. We find that
the rare fluctuations are less important in the running coupling
case as compared to the fixed coupling case.

\section{Fixed coupling case}
\label{sec:BK_eq} The BK equation~\cite{B,K} gives
the evolution with rapidity $Y=\ln(1/x)$ of the scattering
amplitude $S(x_{\bot},y_{\bot},Y)$ of a $q\bar{q}$ dipole with a
target which may be another dipole, a hadron or a nucleus. The
BK equation is a simple equation to deal with the onset of unitarity and to
study parton saturation phenomena at high energies. The analytic
solution to the fixed coupling BK equation for the
$S$-matrix deep in the saturation regime has been derived by Levin
and Tuchin\cite{LT}. This solution agrees with the one derived by
solving the BK equation in the small $S$ limit~\cite{HM}. In this section we will give a simple
derivation of the BK equation and its solution in
the saturation regime.

\subsection{The BK equation}
\label{Sec_Kovchegov_equation}
\begin{figure}[h!]
\setlength{\unitlength}{0.5cm}
\begin{center}
\epsfig{file=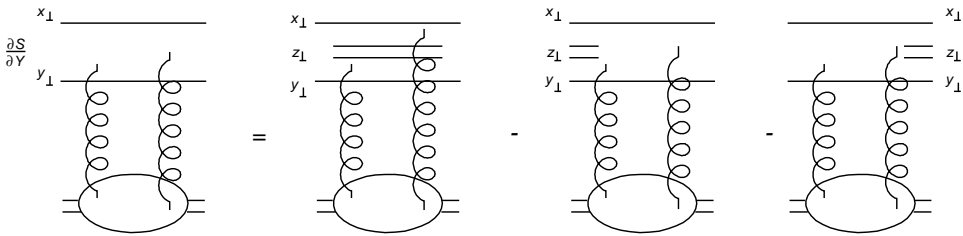, width=14.5cm}
\end{center}
\caption{Diagrams corresponding to terms in the evolution equation~(\ref{eq_kov_1}).}
\label{Fig_Kovchegov}
\end{figure}

In the high-energy scattering of a quark-antiquark dipole on a
target, it is convenient to view the scattering process in a
frame where the dipole is moving along the negative $z$-axis and
the target is moving along the positive $z$-axis. Further we assume
that almost all of the rapidity of the scattering, $Y$, is taken by
the target. We denote the scattering amplitude of a dipole, consisting
of a quark at transverse coordinate $x_{\bot}$ and an antiquark at
transverse coordinate $y_{\bot}$, scattering on a target by
$S(x_{\bot},y_{\bot},Y)$. Now suppose we increase Y by a small
amount $dY$. We wish to know how $S(x_{\bot},y_{\bot},Y)$ changes
with the small amount $dY$. If the rapidity of the dipole is
increased while that of the target is kept fixed, then the dipole
has a probability to emit a gluon due to the change $dY$. We now
calculate the probability for producing this quark-antiquark-gluon
state. In the large $N_c$ limit the quark-antiquark-gluon state
can be viewed as a system of two dipoles -- one of the dipoles
consists of the initial quark and the antiquark part of the gluon
while the other dipole is given by the quark part of the gluon and
the initial antiquark. Using the dipole model the
probability for producing the quark-antiquark-gluon state from
the initial quark-antiquark state is~\cite{M,AM}
\be
dP = \frac{\alpha N_c}{2 \pi^2} d^2z_{\bot} dY
     \frac{(x_{\bot}-y_{\bot})^2}{(x_{\bot}-z_{\bot})^2(z_{\bot}-y_{\bot})^2} \ ,
\ee
where $z_{\bot}$ is the transverse coordinate of the emitted
gluon. The change in the $S$-matrix, $dS$, for a dipole-hadron scattering is given by
multiplying the probability $dP$ with the $S$-matrix
\be
\frac{\partial}{\partial Y} S(x_{\bot}-y_{\bot},Y) =
           \frac{\alpha N_c}{2 \pi^2} \int d^2z_{\bot}
           \frac{(x_{\bot}-y_{\bot})^2}{(x_{\bot}-z_{\bot})^2(z_{\bot}-y_{\bot})^2} \left [
           S^{(2)}(x_{\bot}-z_{\bot},z_{\bot}-y_{\bot},Y) -
           S(x_{\bot}-y_{\bot},Y) \right ] \ ,
\label{eq_kov_1}
\ee
where $S^{(2)}(x_{\bot}-z_{\bot},z_{\bot}-y_{\bot},Y)$ stands for a simultaneous
scattering of the two produced dipoles on the target (see the first diagram on r.h.s of Fig.~\ref{Fig_Kovchegov}).
The last term in~(\ref{eq_kov_1}) describes the scattering of a single dipole on the target because
the gluon is not in the wavefunction of the dipole at the time of the
scattering (see the last two diagrams in Fig.~\ref{Fig_Kovchegov}).

It is hard to directly use Eq.~(\ref{eq_kov_1}) to study problems of parton evolution and parton saturation phenomena at
high density and high energy QCD, since $S^{(2)}$ is not known. Using the mean field approximation
for the gluonic fields in the target
\be
S^{(2)}(x_{\bot}-z_{\bot},z_{\bot}-y_{\bot},Y) =
  S(x_{\bot}-z_{\bot},Y) S(z_{\bot}-y_{\bot},Y) \ ,
\label{fac}
\ee
one gets the Kovchegov equation~\cite{K}
\be
\frac{\partial}{\partial Y} S(x_{\bot}-y_{\bot},Y) =
           \frac{\alpha N_c}{2 \pi^2} \int d^2z_{\bot}
           \frac{(x_{\bot}-y_{\bot})^2}{(x_{\bot}-z_{\bot})^2(z_{\bot}-y_{\bot})^2} \left [
           S(x_{\bot}-z_{\bot},Y) S(z_{\bot}-y_{\bot},Y) -
           S(x_{\bot}-y_{\bot},Y) \right ] \ .
\label{Eq_Kovchegov}
\ee
With $N(x_{\bot}-y_{\bot},Y) = 1-S(x_{\bot}-y_{\bot},Y)$,
another useful version of the Kovchegov equation is obtained
\bea
\frac{\partial}{\partial Y} N(x_{\bot}-y_{\bot},Y) =
           \frac{\alpha N_c}{2 \pi^2} \int d^2z_{\bot}
           \frac{(x_{\bot}-y_{\bot})^2}{(x_{\bot}-z_{\bot})^2(z_{\bot}-y_{\bot})^2}\!\!\!\!&&\!\!\!\left [
           N(x_{\bot}-z_{\bot},Y) + N(z_{\bot}-y_{\bot},Y) \right .
           \nonumber \\
          && \left . \!\!\!-N(x_{\bot}-y_{\bot},Y)- N(x_{\bot}-z_{\bot},Y)
           N(z_{\bot}-y_{\bot},Y) \right ] \ .
\label{Eq_Kovchegov_N}
\eea
Eq.~(\ref{Eq_Kovchegov_N}) has the following probabilistic interpretation: when evolved in rapidity, the initial
quark-antiquark dipole of size $x_{\bot}-y_{\bot}$ decays into two dipoles of size $x_{\bot}-z_{\bot}$
and $z_{\bot}-y_{\bot}$ with the decay probability $ (x_{\bot}-y_{\bot})^2/((x_{\bot}-z_{\bot})^2
(z_{\bot}-y_{\bot})^2)$ which is usually called as BFKL kernel. These two dipoles then interact with the target.
The non-linear term takes into
account a simultaneous interaction of two produced dipoles with the target. The non-linear term prevents the
amplitude from growing boundlessly with rapidity and ensures the unitarity of the scattering amplitude.
When the scattering is weak $N\rightarrow0$, the nonlinear term $N(x_{\bot}-z_{\bot},Y)N(z_{\bot}-y_{\bot},Y)$
can be dropped and the linear equation remaining is the dipole
version~\cite{M} of the
BFKL equation~\cite{KLF,BL}.

%


\subsection{Solution to the BK equation in the saturation regime}
\label{Sec_Solution_Kovchegov}

In the high-energy regime where unitarity corrections become important or
$S(x_{\bot}-y_{\bot},Y)$ is small, Eq.~(\ref{Eq_Kovchegov}) is easier
to use since the quadratic term $S(x_{\bot}-z_{\bot},Y)
S(z_{\bot}-y_{\bot},Y)$ can be neglected in which case one needs only keep the second
term on the r.h.s of~(\ref{Eq_Kovchegov}) giving
\be
\frac{\partial}{\partial Y} S(x_{\bot}-y_{\bot},Y) =
           - \frac{\alpha N_c}{2 \pi^2} \int d^2z_{\bot}
           \frac{(x_{\bot}-y_{\bot})^2}{(x_{\bot}-z_{\bot})^2(z_{\bot}-y_{\bot})^2}
           S(x_{\bot}-y_{\bot},Y)\ .
\label{Eq_Kovchegov_S}
\ee
In the above equation, we have assumed that $S$ is small which holds
only when the dipole size is large compared to $1/Q_s$. Therefore
the lower bound of integration in ~(\ref{Eq_Kovchegov_S}) should
restrict to the regime $(x_{\bot}-y_{\bot})^2\gg 1/Q_s^2$ as well
as to the regime $(x_{\bot}-z_{\bot})^2\gg 1/Q_s^2$,
$(z_{\bot}-y_{\bot})^2\gg 1/Q_s^2$. In the logarithmic regime of
integration one gets
\be
\frac{\partial}{\partial Y} S(x_{\bot}-y_{\bot},Y) =
           - 2\frac{\alpha N_c}{2 \pi^2} \pi\int^{(x_{\bot}-y_{\bot})^2}_{1/Q_S^2}
           d(z_{\bot}-y_{\bot})^2
           \frac{1}{(z_{\bot}-y_{\bot})^2}
           S(x_{\bot}-y_{\bot},Y)\ .
\label{Eq_Kovchegov_S1}
\ee
Note that the factor 2 in the above equation comes from the symmetry of the two regions dominating the integral, either from
$1/Q_S\ll r_1\ll r,~r_2\sim r$ or $1/Q_S\ll r_2\ll r,~r_1\sim r$. Now it is easy to get the solution
to Eq.~(\ref{Eq_Kovchegov_S1})
\be
S(x_{\bot}-y_{\bot},Y)=\exp\left[-\frac{c}{2}\left(\frac{\alpha N_c}{\pi}\right)^2(Y-Y_0)^2\right]S(x_{\bot}-y_{\bot},Y_0)
\label{Sol_Kovchegov},
\ee
where we have used
\be
Q_S^2(Y)=\exp\left[c\frac{\alpha N_c}{\pi}(Y-Y_0)\right]Q_S^2(Y_0)
\label{S_momentum}
\ee
and
\be
Q_S^2(Y_0)(x_{\bot}-y_{\bot})^2=1.
\label{Identity}
\ee

Eq.~(\ref{Sol_Kovchegov}) gives the standard result given in the literature~\cite{HM}. We have gone through such a
detailed ``derivation'' of (\ref{Sol_Kovchegov}) since one of the main purposes of the present paper is to
show how Eq.~(\ref{Sol_Kovchegov}) is modified once running coupling effects and rare fluctuations are included.

\section{Running coupling case}
\label{sec:BK_eqRC}
The BK equation only considers the
resummation of leading logarithmic (LL) $\alpha_s\ln(1/x_{Bj})$
corrections with a fixed coupling constant $\alpha_s$.
The running coupling corrections due to fermion (quark) bubble diagrams, which would bring in a factor of $\alpha_s
N_f$, modify the evolution equation, which is not leading logarithms anymore. Once including $\alpha_s N_f$ corrections,
the obtained contributions have to divided into two parts, running coupling part and the ``subtraction'' part. The first part
has a form as the leading order BK kernel but with the running coupling replacing the fixed coupling and the second part brings
in new structures into the evolution equation.

\begin{figure}[h!]
\setlength{\unitlength}{0.5cm}
\begin{center}
\epsfig{file=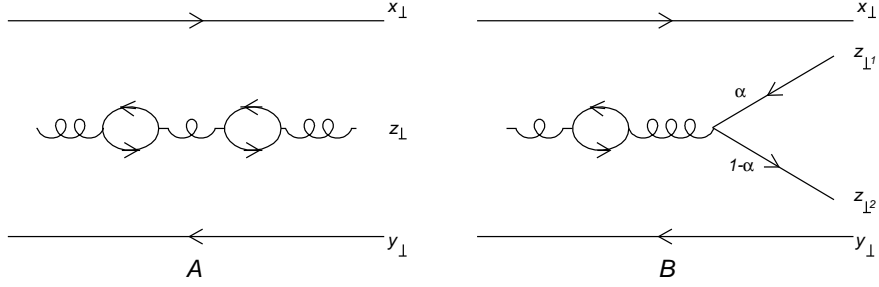, width=12.5cm}
\end{center}
\caption{The higher order diagrams contribution to BK evolution.}
\label{nlo}
\end{figure}

\subsection{Balitsky and Kovchegov-Weigert equations}
%
The evolution equation including higher order corrections reads~\cite{JA}
\be
\frac{\partial S(x_{\bot}-y_{\bot},Y)}{\partial Y}=\mathcal{R}[S]-\mathcal{S}[S]\ .
\label{BK_RC1}
\ee

The first term in r.h.s of (\ref{BK_RC1}), $\mathcal{R}$, which is referred to as the 'running coupling' contribution and
resums all power of $\alpha_s N_f$ corrections
to the evolution. The $\mathcal{R}$ has a form as the leading order one but with modified kernel which includes all
effects of the running coupling
\be
  \mathcal{R}\left[S(x_{\bot}-y_{\bot},Y)\right]= \int\,d^2 z_{\bot}
  \,\tilde{K}(x_{\bot},y_{\bot}, z_{\bot})
  \left[S(x_{\bot}-z_{\bot}Y)\,S(z_{\bot}-y_{\bot},Y)-S(x_{\bot}-y_{\bot},Y)\right]\ .
\label{run}
\ee
The BK kernel is modified because the propagator of the emitted gluon in the original parent dipole is now dressed with
quark loops in contrast to leading order or fixed coupling one. This modifies the emission probability of the gluon
but doesn't change the leading order interaction terms (see Fig.2A).

Using $N(x_{\bot}-y_{\bot},Y)=1-S(x_{\bot}-y_{\bot},Y)$, another
useful version of~(\ref{run}) is:
\bea
  \mathcal{R}\left[N(x_{\bot}-y_{\bot},Y)\right]= \int\,d^2 z_{\bot}
  \,\tilde{K}(x_{\bot},y_{\bot}, z_{\bot})
  \left [
           N(x_{\bot}-z_{\bot},Y) + N(z_{\bot}-y_{\bot},Y) \right .
           \nonumber \\
           \left . ~~~~~~~~~~-N(x_{\bot}-y_{\bot},Y)- N(x_{\bot}-z_{\bot},Y)
           N(z_{\bot}-y_{\bot},Y) \right ] \
\label{run1}
\eea
with modified kernel $\tilde{K}(x_{\bot},y_{\bot}, z_{\bot})$
which has two kinds of expressions since two different separation
schemes of running coupling and subtraction have been used
in~\cite{Bnlo,KW}(see~\cite{JA} for more discussions on separation
schemes). Balitsky took the transverse coordinate of either the
quark at $z_{{\bot}1}$ or the antiquark at $z_{{\bot}2}$ to be the
subtraction point. He got the kernel of the running coupling
contribution as follows~\cite{Bnlo}
\be
  \tilde{K}^{\text{Bal}}({\bf r},{\bf r}_1,{\bf r}_2)=\frac{N_c\,\alpha_s(r^2)}{2\pi^2}
  \left[\frac{r^2}{r_1^2\,r_2^2}+
    \frac{1}{r_1^2}\left(\frac{\alpha_s(r_1^2)}{\alpha_s(r_2^2)}-1\right)+
    \frac{1}{r_2^2}\left(\frac{\alpha_s(r_2^2)}{\alpha_s(r_1^2)}-1\right)
  \right]\,.
\label{kbal}
\ee
Here we introduce the notation ${\bf r}=x_{\bot}-y_{\bot}$, ${\bf
r}_1=x_{\bot}-z_{\bot}$ and ${\bf r}_2=z_{\bot}-y_{\bot}$ for the
sizes of parent and of the new daughter dipoles produced by the
evolution. On the other hand, in the subtraction scheme proposed
by Kovchegov-Weigert the subtraction point is fixed at the
transverse coordinate of the gluon at $z_{\bot}=\eta
z_{{\bot}1}+(1-\eta)z_{{\bot}2}$ in which $\eta$ is the
longitudinal momentum fraction of gluon carried by quark. They got
the modified kernel of the running coupling
contribution~\cite{KW}:
\be
  \tilde{K}^{\text{KW}}({\bf r},{\bf r}_1,{\bf r}_2)=\frac{N_c}{2\pi^2}\left[
    \alpha_s(r_1^2)\frac{1}{r_1^2}-
    2\,\frac{\alpha_s(r_1^2)\,\alpha_s(r_2^2)}{\alpha_s(R^2)}\,\frac{
      {\bf r}_1\cdot {\bf r}_2}{r_1^2\,r_2^2}+
    \alpha_s(r_2^2)\frac{1}{r_2^2} \right]\,
\label{kkw}
\ee
with
\be
R^2({\bf r},{\bf r}_1,{\bf r}_2)=r_1\,r_2\left(\frac{r_2}{r_1}\right)^
{\frac{r_1^2+r_2^2}{r_1^2-r_2^2}-2\,\frac{r_1^2\,r_2^2}{
      {\bf r}_1\cdot{\bf r}_2}\frac{1}{r_1^2-r_2^2}}\,.
\label{r}
\ee

The second term in r.h.s of~(\ref{BK_RC1}), $\mathcal{S}$, which is referred to as 'subtraction' contribution, is given by
\be
  {\mathcal S} [S] \, = \, \alpha_{\mu}^2 \, \int d^2 z_{{\bot}1} \, d^2 z_{{\bot}2} \,
  K_{\oone} (x_{\bot}, y_{\bot} ; z_{{\bot}1}, z_{{\bot}2}) \, [ S (x_{\bot}-
   w_{\bot}, Y) \, S (w_{\bot}-y_{\bot}, Y) - S (x_{\bot}-
  z_{{\bot}1}, Y) \, S (z_{{\bot}2}-y_{\bot}, Y)]
\label{sub_full}
\ee
with $\alpha_{\mu}$ the bare coupling. The interaction structures are modified in the above equation since the quark-antiquark pair
is added to the evolved wave function (see Fig.2B).
The ${\cal K}_{\oone}(x_{{\bot}m}, x_{{\bot}n} ; z_{{\bot}1},
z_{{\bot}2})$ is a resummed JIMWLK kernel which can be found
in~\cite{JA}
\be
  K_{\oone} (x_{\bot}, y_{\bot} ; z_{{\bot}1}, z_{{\bot}2}) \, = \, C_F
  \, \sum_{m,n = 0}^1 \, (-1)^{m+n} \, {\cal K}_{\oone}(x_{{\bot}m},
  x_{{\bot}n} ; z_{{\bot}1}, z_{{\bot}2}).
\label{fullBK}
\ee

In terms of Balitsky' subtraction scheme one substitutes $w_{\bot}=z_{{\bot}1}$ or $w_{\bot}=z_{{\bot}2}$ in Eq.~(\ref{sub_full})
and gets the subtraction term
\be
  {\mathcal S}^{Bal} [S] \, = \, \alpha_{\mu}^2 \, \int d^2 z_{{\bot}1} \, d^2 z_{{\bot}2} \,
  K_{\oone} (x_{\bot}, y_{\bot} ; z_{{\bot}1}, z_{{\bot}2}) \, [ S (x_{\bot}-
   z_{{\bot}1}, Y) \, S (z_{{\bot}1}-y_{\bot}, Y) - S (x_{\bot}-
  z_{{\bot}1}, Y) \, S (z_{{\bot}2}-y_{\bot}, Y)].
\label{sub_Bal1}
\ee
According to Kovchegov-Weigert's subtraction scheme one substitutes $w_{\bot}=z_{\bot}=\eta
z_{{\bot}1}+(1-\eta)z_{{\bot}2}$ in Eq.~(\ref{sub_full}) and gets
\be
  {\mathcal S}^{KW} [S] \, = \, \alpha_{\mu}^2 \, \int d^2 z_{{\bot}1} \, d^2 z_{{\bot}2} \,
  K_{\oone} (x_{\bot}, y_{\bot} ; z_{{\bot}1}, z_{{\bot}2}) \, [ S (x_{\bot}-
   z_{{\bot}}, Y) \, S (z_{{\bot}}-y_{\bot}, Y) - S (x_{\bot}-
  z_{{\bot}1}, Y) \, S (z_{{\bot}2}-y_{\bot}, Y)].
\label{sub_KW1}
\ee

Eq.~(\ref{sub_full}) shows that the ${\mathcal S} [S]$ is of order $\alpha_{\mu}^2$ while ${\mathcal R}[S]$ is of
order $\alpha_s$
and all terms of ${\mathcal S} [S]$ are quadratic
in $S$, $S (x_{\bot}-w_{\bot}, Y) \, S (w_{\bot}-y_{\bot}, Y)$, $S
(x_{\bot}-z_{{\bot}1}, Y) \, S (z_{{\bot}2}- y_{\bot}, Y)$. Thus, for high rapidity and small $S$, the subtraction term
is small as compared to the running coupling term, as also shown numerically in~\cite{JA}. Since this is the kinematic
region in which we are interested in this paper, we hereafter will neglect the subtraction term.
. In this paper we study
the evolution equation in the saturation regime where the
evolution equation including running coupling corrections can be solved
analytically.

\subsection{Solution to Balitsky and Kovchegov-Weigert equations in the saturation regime}
\label{solution_BKRC}
In the saturation regime in which the interaction between partons
is very strong, $S(x_{\bot}-y_{\bot},Y)\rightarrow 0$, and
unitarity corrections become important, the quadratic terms
in~(\ref{BK_RC1}) can be neglected in which case one needs only
keep the second term on the r.h.s of~(\ref{run}). The evolution
equation including running coupling is given by
\be
\frac{\partial S(x_{\bot}-y_{\bot},Y)}{\partial Y}=-\int\,d^2 z_{\bot}
  \,\tilde{K}({\bf r},{\bf r}_1,{\bf r}_2)S(x_{\bot}-y_{\bot},Y)
\label{SKW_sr}
\ee
with modified kernel $\tilde{K}({\bf r},{\bf r}_1,{\bf r}_2)$. In
the saturation region, $rQ_S(Y)\gg 1$, the main contribution to
the integration on the r.h.s of~(\ref{SKW_sr}) comes from either
\be
1/Q_S\ll r_1\ll r;~~~~~~~~~ r_2\sim r
\ee
or
\be
1/Q_S\ll r_2\ll r~~~~~~~~~ r_1\sim r.
\ee
Let us look at one of them, i.e., when $1/Q_S\ll r_1\ll r$, the $r_2$ is approximate equal
to $r$, $r_2\sim r$, the $\tilde{K}^{\text{Bal}}({\bf r},{\bf r}_1,{\bf r}_2)$ becomes
\bea
\tilde{K}^{\text{Bal}}({\bf r},{\bf r}_1,{\bf r}_2)&=& \frac{N_c\alpha_s(r^2)}{2\pi}\left[\frac{1}{r_1^2}\frac{\alpha_s(r_1^2)}
{\alpha_s(r^2)}+\frac{1}{r_1}\left(\frac{\alpha_s(r^2)}{\alpha_s(r_1^2)}-1\right)\right]\nonumber\\
&\approx&\frac{N_c}{2\pi}\frac{\alpha_s(r_1^2)}{r_1^2}
\label{balk}
\eea
and the $\tilde{K}^{\text{KW}}({\bf r},{\bf r}_1,{\bf r}_2)$ has the form as follows
\be
  \tilde{K}^{\text{KW}}({\bf r},{\bf r}_1,{\bf r}_2)=\frac{N_c}{2\pi^2}\left[
    \alpha_s(r_1^2)\frac{1}{r_1^2}-
    2\,\frac{\alpha_s(r_1^2)\,\alpha_s(r_2^2)}{\alpha_s(r_1^2)}\,\frac{
      {\bf r}_1\cdot {\bf r}_2}{r_1^2\,r_2^2}+
    \alpha_s(r_2^2)\frac{1}{r_2^2} \right]\,,
\label{kkw1} \ee here we use $R^2({\bf r},{\bf r}_1,{\bf
r}_2)\approx r_1^2$ which can be obtained via simple calculation
in~(\ref{r}) with condition of $1/Q_S\ll r_1\ll r$ and $r_2\sim
r$. In the $r_1\ll r_2$ limit it is the first term which dominates
Eq.~(\ref{kkw1}) and has the running coupling scale given by the
size of the smaller dipole
\be
  \tilde{K}^{\text{KW}}({\bf r},{\bf r}_1,{\bf r}_2)\approx\frac{N_c}{2\pi^2}
    \alpha_s(r_1^2)\frac{1}{r_1^2}\,.
\label{kkw2}
\ee
We wish to note that the modified Balitsky and Kovchegov-Weigert
kernels including running coupling have the same form in the
saturation regime. It is an interesting outcome which means that the
evolution equation with running coupling corrections is
independent of the choice of transverse coordinate of subtraction
point in the saturation regime. And the modified Balitsky and
Kovchegov-Weigert equations with running coupling corrections are
equivalent to each other in the saturation region. In other words, the
$S$-matrix of the Balitsky and Kovchegov-Weigert equations are exactly
the same in the saturation region.

Now let us put the modified kernel~(\ref{balk}) or~(\ref{kkw2}) into~(\ref{SKW_sr}),
we can get a simplified evolution equation as follows:
\be
\frac{\partial S(r,Y)}{\partial Y}=-\frac{N_c}{2\pi^2}\int_{1/Q_S^2}^{r^2}\,d^2 r_1
  \,\alpha_s(r_1^2)\frac{1}{r_1^2}S(r,Y),
\label{SKW_f}
\ee
with the running coupling at one loop accuracy
\be
\alpha_s(r_1^2)=\frac{\mu}{1+\mu_1\ln\left(\frac{1}{r_1^2\Lambda^2}\right)}
\ee
giving
\be
\frac{\partial S(r,Y)}{\partial Y}= -\frac{N_c\mu}{\pi\mu_1}\left[\ln\left(1+\mu_1\ln\left(\frac{Q_S^2(Y)}{\Lambda^2}\right)\right)-
\ln\left(1+\mu_1\ln\left(\frac{1}{r^2\Lambda^2}\right)\right)\right]S(r,Y)
\ee
whose solution (see also~\cite{Mu}) is
\be
S(r,Y)=e^{-\frac{N_c\mu}{c\pi\mu_1}\left[\ln^2\left(\frac{Q_S^2(Y)}{\Lambda^2}\right)\ln\left(\frac{1+\mu_1\ln\frac{Q_S^2}{\Lambda^2}}{1+\mu_1\ln\frac{1}{r^2\Lambda^2}}-\frac{1}{2}\right)+\frac{\ln\left(\frac{Q_S^2(Y)}{\Lambda^2}\right)}
{\mu_1}-\frac{1}{\mu_1^2}\ln\left(1+\mu_1\ln\frac{Q_S^2}{\Lambda^2}\right)\right]}S(r_0,Y)
\label{solution_running}
\ee
with the saturation momentum
\be
\ln\left(\frac{Q_S^2(Y)}{\Lambda^2}\right)=\sqrt{c(Y-Y_0)}+\textsl{O}(Y^{1/6}).
\label{qsrunning}
\ee
We wish to note that the analytic result for the $S$-matrix including the running
coupling corrections is different as compared to the fixed coupling case. The exponent
in Eq.~(\ref{solution_running}) is
decreasing linearly with rapidity
while the exponent in Eq.~(\ref{Sol_Kovchegov}) is decreasing
quadratically with rapidity, which indicates that the running coupling
slows down the evolution of the scattering amplitude with rapidity.



\section{Effects of rare fluctuations}
\subsection{Fixed coupling case}
At very high energy the typical configuration of a dipole's light-cone wavefunction is a Color Glass Condensate which
is a state having high occupancy for all gluonic levels of momentum less than or equal to saturation momentum $Q_S$.
In the fixed coupling case, the authors of Ref.~\cite{EM} computed the $S$-matrix of two typical
configurations (of condensate type) and of dipole-typical configuration scattering, they found that the typical configurations lead to too small results for the $S$-matrix, being proportional to $\exp\{-c_1 Q_S^2r_0^2/\alpha_s^2\}$ and $\exp\{-\frac{1}{2c}\ln^2(Q_S^2r_0^2)\}$, respectively. $c_1$ and $c$ are constant which are not important for our purpose. Thus they tried to search for configurations which are more rare in the wavefunction but which
dominate very high energy dipole-dipole scattering and lead
to a larger $S$-matrix. They found the reason why the typical configurations have given a small $S$-matrix is that the
typical configurations contain too many gluons at the time of collision, therefore leading to the $S$-matrix is
extremely small. This suggests that the strategy for finding the rare configuration is to minimize the number of gluons by
suppressing the evolution  (see next section for the details of how to obtain the rare configuration). The rare configuration is a state which have no more than
one dipole of size $\kappa r_0$ or larger (with $\kappa$ a constant of order 1 and $r_0$ a size of parent dipole) when the system undertakes BK evolution.
In the center of mass frame, the $S$-matrix is then given by the probability of the rare configurations for each of the parent dipoles partaking in the collision, times the $S$-matrix for the scattering of two dipoles separated by a rapidity gap $Y_0$,
\be
S_Y\approx e^{-\frac{1}{4c}\ln^2(Q_S^2r_0^2)}S_{Y_0}(r_0)
\ee
which is significantly larger than the results coming from the condensate-condensate and dipole-condensate scattering.



%
\begin{figure}[h!]
\setlength{\unitlength}{0.2cm}
\begin{center}
\epsfig{file=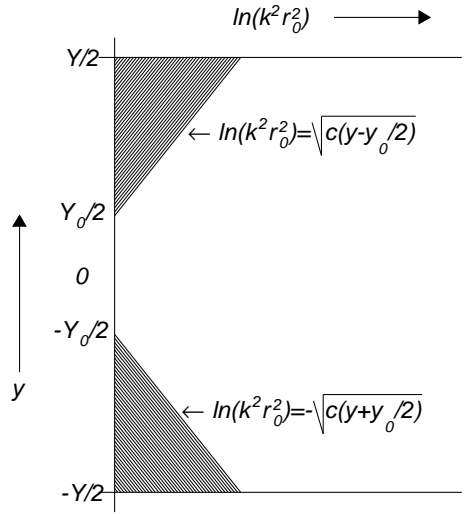, width=10.5cm}
\end{center}
\caption{The configuration in center of mass frame.}
\label{fig2}
\end{figure}
\subsection{Running coupling case}
\label{rarerunning}
Following the framework of Ref.~\cite{EM}, consider the high-energy scattering of dipoles at zero impact
parameter in the center of mass frame where one of dipoles is
left-moving and the other is right-moving. In order to obtain rare configuration, we suppose that the
wavefunction of the right-moving dipole consists only of the parent dipole
with size $r_0$ in the rapidity interval $Y_0/2<y<Y/2$, where
$Y_0$ is the critical value of rapidity for the onset of unitarity
corrections, with the similar requirement on the left-moving
dipole in the rapidity interval $-Y/2<y<-Y_0/2$. In the rapidity
interval $0<y<Y_0/2$ and $-Y_0/2<y<0$ the right-moving and
left-moving dipoles have normal BFKL evolution, respectively.

However, we cann't require that all evolution
of right-moving dipoles are absent in the rapidity interval $Y_0/2<y<Y$. What we can do is to only allow
that evolution which produces very small dipoles, in order to guarantee the system
have no more than one dipole of size $\kappa r_0$ or larger, with
$\kappa$ a constant of order one. And we setup constraints to
suppress the creation of dipoles much smaller than $r_0$ at
rapidities $y>Y_0/2$ to avoid dipoles emitted at intermediate
rapidities evolving into dipoles of size $r_0$ or larger at
rapidity $Y/2$. We require that the gluon emission
from the parent dipoles is forbidden if the gluon has $k_{\bot}$
and $y$ in the shaded triangles of Fig.3. The line
\be
\ln(k_{\bot}r_0^2)=\sqrt{c(y-\frac{Y_0}{2})}
\ee
and a similar line for the lower triangle, is determined by the
requirement that gluons in the right hand side of that
line cann't evolve by normal BFKL evolution into shaded triangles.

Now we compute the probability
of rare configurations $\mathbb{S}(x_{\bot}-y_{\bot},Y-Y_0)$ which has the same
meaning as the survival probability of the parent dipoles after a BFKL evolution
over a rapidities interval $Y-Y_0$~\cite{EM}. This probability decreases with
increasing $Y$ due to gluon emission and the corresponding rate is
the same as the virtual term in~$(\ref{run1})$:
\be
  \frac{\partial}{\partial Y}\mathbb{S}(x_{\bot}-y_{\bot},Y-Y_0)=- \int\,d^2 z_{\bot}
  \,\tilde{K}(x_{\bot},y_{\bot}, z_{\bot})\mathbb{S}(x_{\bot}-y_{\bot},Y-Y_0)
\label{run2}
\ee
whose solution is similar to Eq.~(\ref{solution_running}):
\be
\mathbb{S}(r,Y-Y_0)=e^{-\frac{N_c\mu}{c\pi\mu_1}\left[\ln^2\left(\frac{Q_S^2(Y)}{\Lambda^2}\right)
\ln\left(\frac{1+\mu_1\ln\left(\frac{Q_S^2(Y)}{\Lambda^2}\right)}{1+\mu_1\ln\left(\frac{1}{r^2\Lambda^2}\right)}
-\frac{1}{2}\right)+\frac{\ln\left(\frac{Q_S^2(Y)}{\Lambda^2}\right)}{\mu_1}-\frac{1}{\mu_1^2}\ln\left(1+\mu_1
\ln\left(\frac{Q_S^2(Y)}{\Lambda^2}\right)\right)\right]}.
\label{solution_running1}
\ee
%
%

Let $\mathbb{S}(r_0,(Y-Y_0)/2)$ denote the probability of
a parent dipole not given rise to any emission of gluon in the upper triangle of Fig.3. The $S$-matrix can
be obtained by the product of $\mathbb{S}(r_0,(Y-Y_0)/2)$ for each of
the parent dipoles participating in the scattering, times
$S(r_0,Y_0)$ which is a $S$-matrix for the scattering of two elementary
dipoles. By using
$(\ref{solution_running1})$, one gets:
\bea
S(r,Y)=e^{-\frac{N_c\mu}{c\pi\mu_1}\left[\ln^2\left(\frac{Q_S^2(Y)}{\Lambda^2}\right)
\ln\left(\frac{1+\frac{\mu_1}{\sqrt{2}}\ln\left(\frac{Q_S^2(Y)}{\Lambda^2}\right)}
{1+\mu_1\ln\left(\frac{1}{r^2\Lambda^2}\right)}-\frac{1}{2}\right)+\frac{\sqrt{2}
\ln\left(\frac{Q_S^2(Y)}{\Lambda^2}\right)}{\mu_1}-\frac{2}{\mu_1^2}\ln\left(1+\frac{\mu_1}
{\sqrt{2}}\ln\left(\frac{Q_S^2(Y)}{\Lambda^2}\right)\right)\right]}S(r_0,Y_0)
\label{scms}
\eea
which only brings in very small corrections
to~(\ref{solution_running}) and indicates that the rare fluctuations are less important
in the running coupling case
as compared to the fixed coupling
case~\cite{EM}, where the rare fluctuations are important and the
exponential factor of $S$-matrix in the saturation regime has
twice as large as the result which emerges when fluctuations are
taken into account. We also consider the rare fluctuations  on top of
the running coupling effects in a general frame (please see the Appendix), we find the same result as (\ref{scms}).

\section{Shape of dipole cross section with running coupling}
The authors of Ref.~\cite{DM} computed the scattering
amplitude for $rQ_S\ll 1$ using BFKL evolution and running coupling.
Combining the outcome of Ref.~\cite{DM} and our result
(\ref{solution_running}) which is valid for $rQ_S\gg 1$, the shape of dipole cross
section with running coupling reads:

\begin{displaymath}
N(r,Y)=\left\{\begin{array}{ll}
               \left(\frac{Q^2}{Q_S^2}\right)^{-(1-\lambda_0)}\left[\ln\left(\frac{Q^2}{Q_S^2}
               \right)+\frac{1}{1+\lambda_0}\right]N_0 ~~~~~~~~~~~~~~~~~~~~~~~~~~~~~~~~~~~~~~~~~~~~~~~~~~~~~~~~~~~~~rQ_S\leq1~, \\
            1-e^{-\frac{N_c\mu}{c\pi\mu_1}\left[\ln^2\left(\frac{Q_S^2(Y)}{\Lambda^2}\right)\ln
            \left(\frac{1+\mu_1\ln\left(\frac{Q_S^2(Y)}{\Lambda^2}\right)}{1+\mu_1\ln\left(\frac{1}{r^2\Lambda^2}\right)}
            -\frac{1}{2}\right)+\frac{\ln\left(\frac{Q_S^2(Y)}{\Lambda^2}\right)}{\mu_1}-\frac{1}{\mu_1^2}\ln\left(1+\mu_1\ln
            \left(\frac{Q_S^2(Y)}{\Lambda^2}\right)\right)\right]}S_0 ~~~~rQ_S>1~,
              \end{array}\right.
\end{displaymath}
where $\lambda_0$ is the solution to
$\chi^{'}(\lambda_0)(1-\lambda_0)=-\chi(\lambda_0)$ with $\chi$ the usual
BFKL eigenvalue function, $N_0$ is a constant but with no control
at this moment and $Q_S$ is the saturation momentum including running
coupling corrections.


\begin{acknowledgments}
I would like to thank my advisor Arif.~Shoshi for
suggesting this work and numerous stimulating discussions. Without his patient guidance,
this work would not be possible.
The author acknowledge financial support by the Deutsche Forschungsgemeinschaft under
contract Sh 92/2-1.
\end{acknowledgments}

%
\appendix
\section{Rare fluctuations in a general frame}

Consider a high energy scattering of a right-moving dipole of size
$r_0$ and rapidity $Y-Y_2$ on a left-moving dipole of size $r_1$
and rapidity $-Y_2$ in an arbitrary frame. The frame and
scattering picture are illustrated in Fig.4, where $Y_0$ is a
rapidity gap between two dipoles. For later convenience, we
require that $Y_2\leq\frac{1}{2}(Y-Y_0)$. We require that no
additional dipoles can be created from the gluon emission of
left-moving dipole $r_1$ which would have a strong interaction
with the right-moving dipoles. The dipoles which would have such
strong interactions would be of size $r\geq 1/Q_S$ at the scattering time.
So we should suppress the emission
of those dipoles which could become of size $1/Q_S$ or larger
after a normal evolution over the rapidity interval $-Y_2<y<0$.
For the right-moving dipole $r_0$, we suppress evolution over its
$Y-Y_2-(Y_1+Y_0)$ with the region of suppression given by the
upper shaded triangle of Fig.4. The line
\be
\ln(k^2r_0^2)=\sqrt{c(y-Y_1-Y_0)}
\ee
and a similar line for the lower triangle, is determined by the
requirement that gluons locating in the right hand side of that
line cann't evolve by normal BFKL evolution into shaded triangles.
We will determine $Y_1$ by maximizing the $S$-matrix later.
The unshaded triangle, whose rapidity values go from
$0$ to $Y_1$, is a saturation region where the dipole $r_0$ has
evolved into a Color Glass Condensate.

After we have a clear scattering picture of dipoles, the $S$-matrix can be evaluated at hand
\be
S(r_0,r_1,Y)=\mathbb{S}_R(r_0,Y-Y_0-Y_1-Y_2)\mathcal{S}(r_0,r_1,Y_0+Y_1)\mathbb{S}_L(r_1,Y_2)
\label{stot}
\ee
with $\mathcal{S}(r_0,r_1,Y_0+Y_1)$ is the $S$-matrix for
scattering of a elementary dipole $r_1$ on a Color Glass
Condensate state which is evolved from dipole $r_0$ and
$\mathbb{S}_R(r_0,Y-Y_0-Y_1-Y_2)$ and $\mathbb{S}_L(r_1,Y_2)$ are
the suppression factor from the no emission requirement for two
dipoles, which are given in terms of the area of the upper and
lower shaded regions of Fig.4. After using
(\ref{solution_running2}), one obtains
\be
\mathbb{S}_R(r_0,Y-Y_0-Y_1-Y_2)=e^{-\frac{N_c\mu}{c\pi\mu_1}\left[c\ln\left(\frac{1+\mu_1\sqrt{c(Y-Y_2-Y_1-Y_0)}}
{1+\mu_1\ln\left(\frac{1}{r^2\Lambda^2}\right)}-\frac{1}{2}\right)(Y-Y_2-Y_1-Y_0)+\frac{\sqrt{c(Y-Y_2-Y_1-Y_0)}}
{\mu_1}-\frac{1}{\mu_1^2}\ln\left(1+\mu_1\sqrt{c(Y-Y_2-Y_1-Y_0)}\right)\right]}
\label{sr}
\ee
and
\bea
\mathbb{S}_L(r_1,Y_2)&=&\exp\left[-\frac{N_c\mu}{c\pi\mu_1}\left[c\ln\left(\frac{1+\mu_1\sqrt{c(Y_1+Y_2)}}
{1+\mu_1\ln\left(\frac{1}{r^2\Lambda^2}\right)}-\frac{1}{2}\right)(Y_1+Y_2)+\frac{\sqrt{c(Y_1+Y_2)}}
{\mu_1}-\frac{1}{\mu_1^2}\ln\left(1+\mu_1\sqrt{c(Y_1+Y_2)}\right) \right.\right.\nonumber \\
~~~~~~~~~~~~~&-&\left.\left.c\ln\left(\frac{1+\mu_1\sqrt{cY_1}}{1+\mu_1\ln\left(\frac{1}{r^2\Lambda^2}\right)}
-\frac{1}{2}\right)Y_1-\frac{\sqrt{cY_1}}{\mu_1}+\frac{1}{\mu_1^2}\ln\left(1+\mu_1\sqrt{cY_1}\right)\right]\right].
\label{sl}
\eea

The $\mathcal{S}$ can be computed by using the BK equation with
running coupling corrections since the BK equation with running
coupling corrections describes correctly the scattering of an
elementary dipole on a Color Glass Condensate. By using
(\ref{solution_running}), one gets
\be
\mathcal{S}(r_0,r_1,Y_0+Y_1)=e^{-\frac{N_c\mu}{c\pi\mu_1}\left[c\ln\left(\frac{1+\mu_1\sqrt{cY_1}}
{1+\mu_1\ln\left(\frac{1}{r^2\Lambda^2}\right)}-\frac{1}{2}\right)Y_1+\frac{\sqrt{cY_1}}{\mu_1}-\frac{1}
{\mu_1^2}\ln\left(1+\mu_1\sqrt{cY_1}\right)\right]}S(r_0,Y_0).
\label{ss}
\ee
\begin{figure}[h!]
\setlength{\unitlength}{0.5cm}
\begin{center}
\epsfig{file=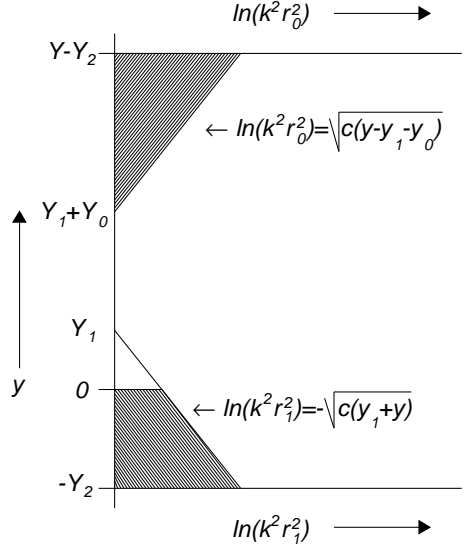, width=10.5cm}
\end{center}
\caption{The configuration in a general frame.}
\label{fig3}
\end{figure}

Substituting (\ref{sr}), (\ref{sl}) and (\ref{ss}) into (\ref{stot}), one obtains:
\bea
S(r_0,r_1,Y)&=&\exp\left[-\frac{N_c\mu}{c\pi\mu_1}\left[c\ln\left(\frac{1+\mu_1\sqrt{c(Y-Y_2-Y_1-Y_0)}}
{1+\mu_1\ln\left(\frac{1}{r^2\Lambda^2}\right)}-\frac{1}{2}\right)(Y-Y_2-Y_1-Y_0)+\frac{\sqrt{c(Y-Y_2-Y_1-Y_0)}}{\mu_1}\right.\right.\nonumber\\
&-&\left.\left.\frac{1}{\mu_1^2}\ln\left(1+\mu_1\sqrt{c(Y-Y_2-Y_1-Y_0)}\right)
+c\ln\left(\frac{1+\mu_1\sqrt{c(Y_1+Y_2)}}{1+\mu_1\ln\left(\frac{1}{r^2\Lambda^2}\right)}-\frac{1}{2}\right)
(Y_1+Y_2)+\frac{\sqrt{c(Y_1+Y_2)}}{\mu_1}\right.\right.\nonumber\\
&-&\left.\left.\frac{1}{\mu_1^2}\ln\left(1+\mu_1\sqrt{c(Y_1+Y_2)}\right)\right]\right]S(r_0,Y_0).
\label{smatrix1}
\eea
which connects to a set of configurations of the wavefunction
described by rapidity $Y_1$. The $S$-matrix is determined by the
values of $Y_1$ which maximizes the r.h.s of Eq.~(\ref{smatrix1})
or equivalently minimizes the exponent of the r.h.s of
Eq.~(\ref{smatrix1}). We obtain
\be
Y_1=\frac{1}{2}(Y-Y_0)-Y_2.
\ee
Take this $Y_1$ into~(\ref{smatrix1}), finally the $S$-matrix is:
\bea
S(r,Y)=e^{-\frac{N_c\mu}{c\pi\mu_1}\left[\ln^2\left(\frac{Q_S^2(Y)}{\Lambda^2}\right)\ln\left(\frac{1+\frac{\mu_1}{\sqrt{2}}
\ln\left(\frac{Q_S^2(Y)}{\Lambda^2}\right)}{1+\mu_1\ln\left(\frac{1}{r^2\Lambda^2}\right)}-\frac{1}{2}\right)+\frac{\sqrt{2}
\ln\left(\frac{Q_S^2(Y)}{\Lambda^2}\right)}{\mu_1}-\frac{2}{\mu_1^2}\ln\left(1+\frac{\mu_1}{\sqrt{2}}\ln\left(\frac{Q_S^2(Y)}
{\Lambda^2}\right)\right)\right]}
\label{arbf}
\eea
which is exactly the same as the corresponding result~(\ref{scms}) in the center of mass frame.


\end{document}